\documentclass[manuscript,screen,anonymous=false]{acmart}
\usepackage{amsmath}
\usepackage{bbm}
\usepackage{booktabs}
\usepackage{multirow}

\usepackage{caption}

\DeclareMathOperator*{\argmax}{arg\,max}

\usepackage{bbm}
\usepackage{figsize}
\usepackage{todonotes}

\newif\ifworkinprogress
\workinprogresstrue

\ifworkinprogress
    \newcommand{\HA}[1]{\textcolor{purple}{[Himan] #1}}
    \newcommand{\MM}[1]{\textcolor{red}{[Masoud] #1}}
  \newcommand{\RB}[1]{\textcolor{blue}{[Robin] #1}}
  \newcommand{\BM}[1]{\textcolor{orange}{[Bamshad] #1}}

\else
  \newcommand{\HA}[1]{}
  \newcommand{\MM}[1]{}
  \newcommand{\RB}[1]{}
  \newcommand{\BM}[1]{}
\fi

\AtBeginDocument{%
  \providecommand\BibTeX{{%
    \normalfont B\kern-0.5em{\scshape i\kern-0.25em b}\kern-0.8em\TeX}}}
\renewcommand\footnotetextcopyrightpermission[1]{}

\begin{document}




\title{Addressing the Multistakeholder Impact of Popularity Bias in Recommendation Through Calibration}



\author{Himan Abdollahpouri}
\affiliation{%
  \institution{University of Colorado Boulder}
  \city{Boulder}
  \country{USA}
}
 \email{himan.abdollahpouri@colorado.edu}

\author{Masoud Mansoury}
\affiliation{%
  \institution{Eindhoven University of Technology}
  \city{Eindhoven}
  \country{Netherlands}
  }
  \email{m.mansoury@tue.nl}

\author{Robin Burke}
\affiliation{%
  \institution{University of Colorado Boulder}
  \city{Boulder}
  \country{USA}
  }
  \email{robin.burke@colorado.edu}

\author{Bamshad Mobasher}
\affiliation{%
  \institution{DePaul University}
  \city{Chicago}
  \country{USA}
  }
  \email{mobasher@cs.depaul.edu}

\renewcommand{\shortauthors}{Abdollahpouri et al.}

\begin{abstract}
Popularity bias is a well-known phenomenon in recommender systems: popular items are recommended even more frequently than their popularity would warrant, amplifying long-tail effects already present in many recommendation domains. Prior research has examined various approaches for mitigating popularity bias and enhancing the recommendation of long-tail items overall. The effectiveness of these approaches, however, has not been assessed in multistakeholder environments where in addition to the users who receive the recommendations, the utility of the suppliers of the recommended items should also be considered. In this paper, we propose the concept of popularity calibration which measures the match between the popularity distribution of items in a user's profile and that of the recommended items. We also develop an algorithm that optimizes this metric. In addition, we demonstrate that existing evaluation metrics for popularity bias do not reflect the performance of the algorithms when it is measured from the perspective of different stakeholders. Using music and movie datasets, we empirically show that our approach outperforms the existing state-of-the-art approaches in addressing popularity bias by calibrating the recommendations to users' preferences. We also show that our proposed algorithm has a secondary effect of improving supplier fairness. 
\end{abstract}




%


\maketitle

\section{Introduction}\label{intro}

Historically, recommendation algorithms were developed to maximize the accuracy of the delivered recommendations to the users. However, as other researchers have noted, there are other important characteristics of recommendations that must be considered, including diversity, serendipity,  novelty~\cite{Vargas:2011:RRN:2043932.2043955,ge2010beyond,castells2011novelty}, and fairness \cite{yao2017beyond}. These characteristics can have enormous impacts on the utility of the recommendations across all system stakeholders \cite{abdollahpourimultistakeholder2020}. 

In this paper, we focus on the problem of \textit{popularity bias}, the tendency of recommender systems to favor a small set of popular items in their recommendations, even more than their popularity would warrant, and to disfavor items that lie outside of this set, even when these items are preferred by a significant number of interested users~\cite{park2008long,steck2011item,jannach2015recommenders}.

\begin{figure*}[tb]
\centering
\SetFigLayout{2}{1}
 \subfigure[Impact on Item Groups]{
 \includegraphics[width=5.8in]{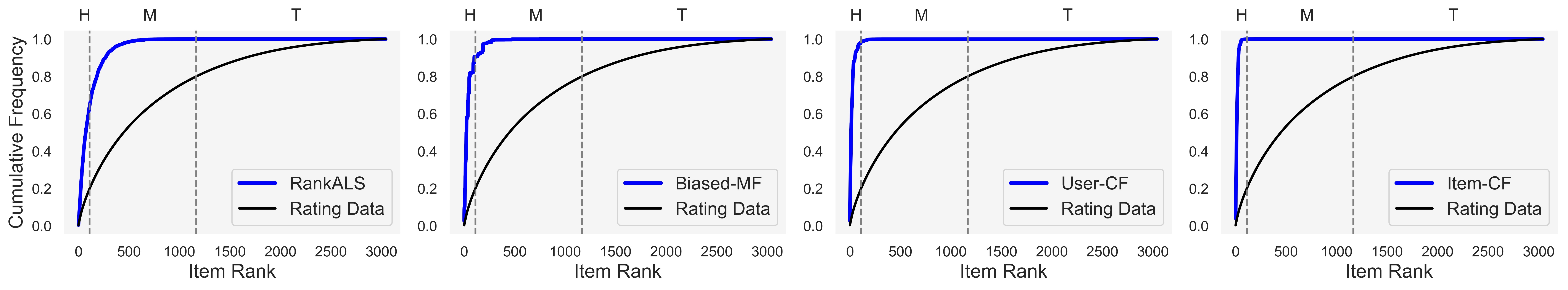}\label{amplify_item_groups}}
  \hfill
  
 \subfigure[Impact on User Groups]{\includegraphics[width=5.8in]{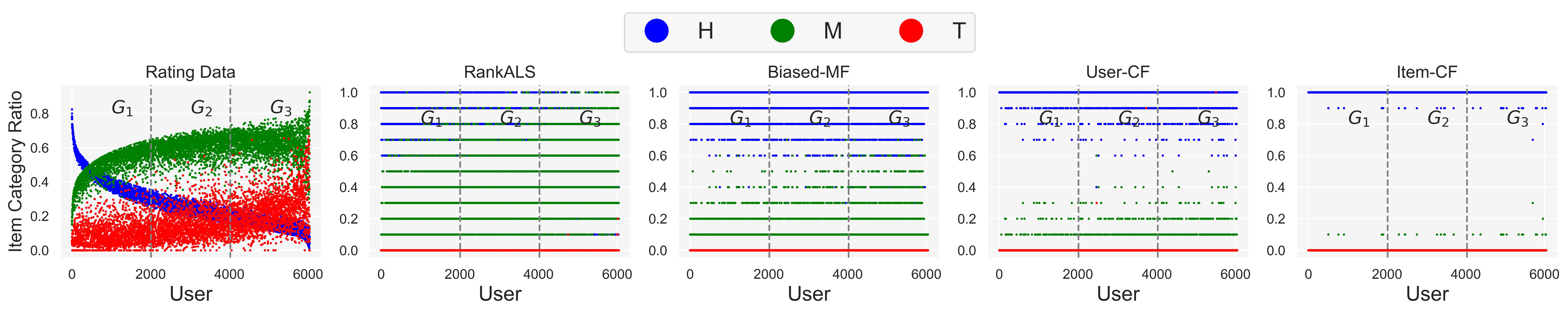}\label{amplify_user_groups}}
  \hfill
  
   \subfigure[Impact on Supplier Groups]{\includegraphics[width=5.8in]{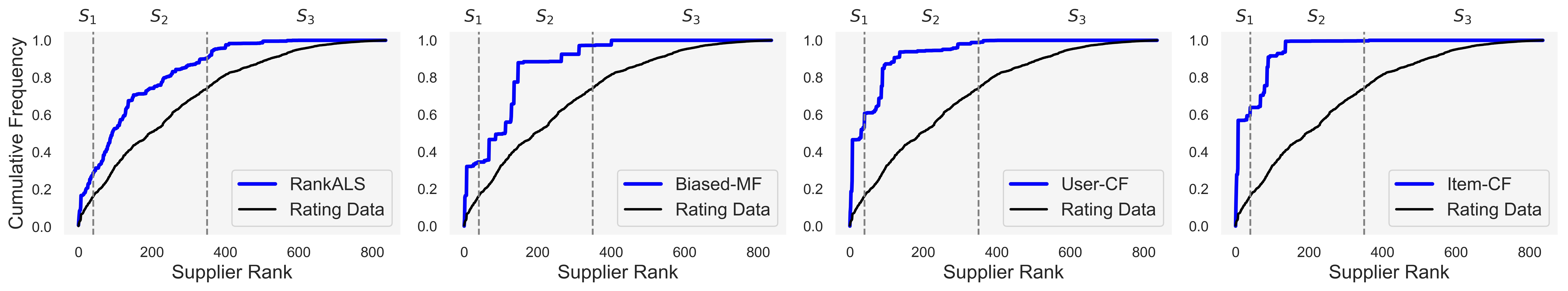}\label{amplify_supplier_groups}}
  \hfill

\caption{Multistakeholder impact of popularity bias amplification \cite{abdollahpouri2020multi}} \label{pop_amplify}
\end{figure*} 

Consider the distributions shown in Figure \ref{amplify_item_groups}. These four plots contrast item popularity and recommendation popularity for four well-known recommendation algorithms (\textit{RankALS} \cite{takacs2012alternating}, \textit{Biased Matrix Factorization (Biased-MF)} \cite{koren2009matrix}, \textit{User-based Collaborative Filtering (User-CF)} \cite{aggarwal2016neighborhood}, and \textit{Item-based Collaborative Filtering (Item-CF)} \cite{sarwar2001item}) using the MovieLens 1M \cite{movielens} data set (See section \ref{data_section} for more details on this dataset). The x-axis indicates the rank of each item when sorted from most popular to least popular. The black curve shows the cumulative frequency of the ratings for different items at a given rank. As we can see, a few highly ranked items dominate the entire rating history. For instance, only 111 items (less than 3\%) take up more than 20\% of the ratings. In many consumer taste domains, where recommender systems are commonly deployed, the same highly-skewed distribution of user interest is seen. A music catalog might contain artists whose songs have been played millions of times (Beyonc\'{e}, Ed Sheeran) and others whose songs have a much smaller audience (Iranian musician Kayhan Kalhor, for example). These few popular items are referred to as the \textit{short-head} (shown by $H$ in the plots) in the literature which take up roughly 20\% of the ratings according to the Pareto Principle \cite{sanders1987pareto}. The rest of the items in a long-tail distribution are usually divided into two other parts \cite{celma2008hits}: Tail items ($T$) are the larger number of less popular items which collectively take up roughly 20\% of the ratings\footnote{Tail items may also be new items that have yet to find their audience and will eventually become popular. In this way, popularity bias is related to the well-known item cold-start problem in recommendation.}, and \textit{Mid} items ($M$) include a larger number of items in between that receive around 60\% of the ratings, collectively. These three item groups are shown on the top of each plot partitioning the items based on their popularity into the most popular items ($H$), items with medium popularity ($M$), and less popular items ($T$). 

The blue curves in each plot show popularity bias at work across the four algorithms. In the most extreme case (\textit{Item-CF}), almost no item beyond rank 111 is recommended. The head of the distribution constituting less than 3\% of the total items take up almost 100\% of the recommendations given to the users. In \textit{User-CF} this number is 99\%. The other algorithms are only slightly better in this regard, with the \textit{H} items taking up more than 64\% and 74\% of the recommendations in \textit{RankALS} and \textit{Biased-MF}, respectively. 

These plots provide an illustration of popularity bias in recommendation. Users' rating profiles exhibit skewed popularity distributions and recommendation algorithms tend to amplify this bias, yielding a ``rich get richer'' dynamic. However, we argue that this aggregate view of recommendation frequency is an incomplete picture of the impact of popularity bias: our goal is to provide a more textured view informed by recent work in multistakeholder recommendation \cite{abdollahpourimultistakeholder2020}. Authors in \cite{abdollahpourimultistakeholder2020} distinguish among the main stakeholders in recommender systems: 1) users (those who receive the recommendations) and 2) suppliers (those who supply or, otherwise, stand behind the recommended items). The popularity bias impacts both of these stakeholders as we see in Figure ~\ref{amplify_user_groups} (impact on users) and ~\ref{amplify_supplier_groups} (impact on suppliers).


The first plot in Figure ~\ref{amplify_user_groups} shows the ratio of rated items for the three item groups $H$, $M$, and $T$ in the profiles of different users in the MovieLens 1M dataset. The interest of each user towards the three item groups is measured and the users are sorted from the highest interest towards popular items to the lowest based on the ratio of different item groups in their profile. Users are first sorted based on interest towards $H$ items and if there is a tie the interest towards $M$ will be considered and if still there is a tie, interest towards $T$ will be compared. Sorted users are divided into three equal-sized bins $G=\{G_1, G_2, G_3\}$ from most popularity-focused ($G_1$) to least ($G_3$). The y-axis shows the proportion of each user's profile devoted to different item groups. The narrow blue band shows the proportion of each users profile that consists of popular items ($H$), and its monotone decrease reflects the way the users are ranked. Note, however, that all user groups, even $G_1$ have rated many items from the \textit{Mid} (green) and \textit{Tail} (red) parts of the distribution, and this makes sense: there are only a few really popular movies and even the most blockbuster-focused viewer will eventually run out of them. 

Other plots in Figure~\ref{amplify_user_groups} also show users ordered by their popularity interest, but now the y-axis shows the proportion of recommended items delivered by different algorithms\footnote{The plots are ``banded'' because recommendation lists are of size 10 and there are only 10 possible ratio values for each item group.}. The difference with the original user profiles in rating data is stark, especially in the case of \textit{Item-CF} and \textit{User-CF} where the users' profiles are rich in diverse item groups, the generated recommendations are much less so. \textit{Tail} items do not appear at all, evidenced by the red line at the bottom of the plots. In \textit{Item-CF} almost 100\% of the recommendations are from the \textit{Head} category, even for the users with the most niche-oriented profiles. The ``head-focused'' $G_1$ users are getting recommendations that are not well-matched to their interests in terms of item popularity, but the $G_3$ ``tail-focused'' users are quite poorly served, getting a steady diet of popular movies in which they are less interested with none of their long-tail interests served. Thus, we see that popularity bias has a differential impact across the user base: all are affected but some much more severely than others. 

Multistakeholder recommendation expands our view of those impacted by recommendations to also include \textit{suppliers}, individuals or entities that stand behind the recommended items. In this paper, we have considered the director of each movie as the supplier for that movie. Figure~\ref{amplify_supplier_groups} examines popularity bias from the supplier-side of the recommendation interaction: the directors whose movies are being recommended. The plots here are comparable to Figure~\ref{amplify_item_groups}, but here we look at the popularity of the movie directors in the data, also finding a skewed distribution indicating movies from few popular directors have taken up large portion of the ratings. Similar to item groups, we have defined three supplier groups based on their popularity $S=\{S_1, S_2, S_3\}$: $S_1$ represents few popular suppliers whose items take up 20\% of the ratings, $S_2$ are larger number of suppliers with medium popularity whose items take up around 60\% of the ratings, and $S_3$ are the less popular suppliers whose items get 20\% of the ratings\footnote{The popularity distribution of items and suppliers has a strong long-tail shape. Therefore, in order to create groups with different popularity values we cannot force the groups to have equal sizes as this would lead to groups with medium and extreme popularity to have very close average popularity.}.

Looking at the results returned by the recommendation algorithms, we see that popularity bias affects the suppliers of the recommendation interaction as well. Strikingly, in \textit{Item-CF}, movies from just 3 popular directors in $S_1$ (less than 0.4\%) take up 50\% of recommendations produced, while the items from $S_3$ are never recommended. The pattern repeats across the other algorithms.

What this analysis demonstrates is that popularity bias is a multistakeholder phenomenon, that entities on different sides of the recommender system are impacted by it (sometimes quite severely) and that any evaluation of the system's performance must capture the perspective of different stakeholders. In this paper, we propose an algorithmic approach based on the novel notion of \textit{popularity calibration} that provides a comprehensive solution to the problem of popularity bias by addressing the needs of multiple stakeholders. A recommendation list is calibrated based on popularity when the range of items it covers matches the user's profile in terms of item popularity. For example, if a user has 20\% $H$ items, 40\% $M$ items and 40\% $T$ items in her profile, we call the recommendations to be calibrated when the ratio of each item group in the recommended list is consistent with the aforementioned ratios. We show that our approach, while yielding more calibrated recommendations for users, also results in fairer recommendations from the suppliers' perspective. 

It is worth noting that calibrating the recommendations in terms of popularity does not necessarily guaranty that the recommended items will be matched to the users' overall preferences. For instance, a documentary movie and a film noir movie might be both non-popular but a user might be interested in one while disliking the other one. With that said, the popularity calibration method we propose in this paper does not aim to calibrate the recommendation lists in terms of content but rather only based on the popularity of the items. 

Our contributions are as follows:

\begin{itemize}
    \item We characterize popularity bias in multistakeholder terms as a problem impacting the entire recommendation ecosystem, and introduce metrics for quantifying its impact.
            
    \item We show that some of the existing metrics in the literature to evaluate popularity bias mitigation often hide important information about how a certain algorithm controls popularity bias for different stakeholders. 

    \item We demonstrate the weaknesses of state-of-the-art re-ranking algorithms when examined from a multistakeholder perspective.

    \item We propose a new simple, yet effective, algorithm for enhancing popularity calibration using re-ranking, a method that can be applied to the output of any recommendation algorithm.
    
    \item Using music and movie datasets, we show that our method outperforms state-of-the-art baselines in mitigating the impact of popularity bias when the needs of multiple stakeholders are considered.
\end{itemize}

\begin{figure}
    \centering
    \includegraphics[width=4in]{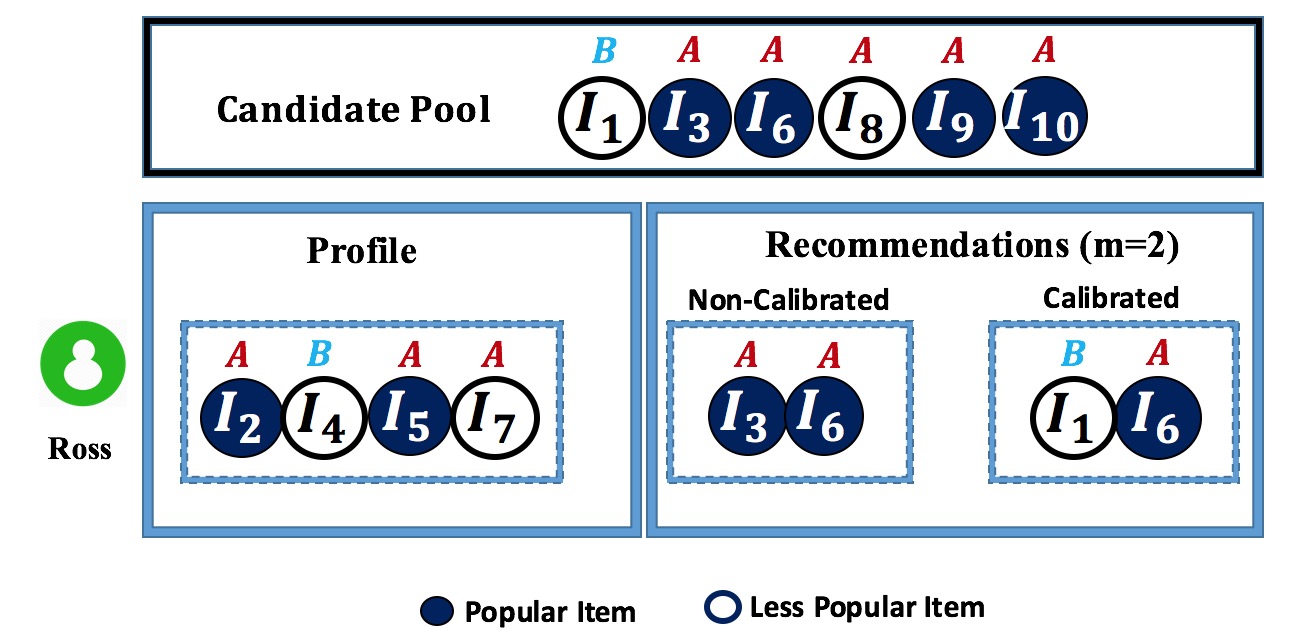}
    \caption{Popularity calibration example}
    \label{fig:rec_graph2}
\end{figure}

\section{Motivating Example}

Figure ~\ref{fig:rec_graph2} shows a user whose name is Ross\footnote{Ross is the name of the first author's favorite character in Friends.}. Out of all the items in a movie recommender's catalog, Ross has interacted with (or liked) four items $I_2$, $I_4$, $I_5$, and $I_7$. Suppose dark circles are popular items ($H$) and white ones are less popular items (either $M$ or $T$). That means his profile consists of an equal ratio of popular versus non-popular items (50\% each), and we assume that the contents of his profile represents the scope of his interest in movies. In addition, each item is provided by one of the suppliers $A$ or $B$. $A$ (red color) who is a popular supplier (based on the average popularity of their items) owns $I_2$, $I_3$, $I_5$, $I_6$, $I_7$, $I_8$, $I_9$ and $I_{10}$. $B$ who is a less popular supplier owns $I_1$ and $I_4$. Note that a popular supplier could also own some less popular items ($A$ owns $I_8$). Since Ross is equally interested in popular and less popular items based on his profile, a well-calibrated recommender should seek to deliver the same ratio in his recommendations, reflecting the diversity of interests that he has shared with the system. Assuming the size of the recommendation set is 2 (for illustration purpose), he should get one popular and one less popular recommendation. A recommendation algorithm influenced by popularity bias would be more likely to generate a recommendation list containing only popular items: ``Non-Calibrated'' in the figure. With this set of recommendations, the supplier $B$ is completely out of the picture and has received zero exposure even though Ross's profile does show interest in less popular movies. The calibrated recommendations, on the right side, shows preferred situation where the user has received recommendations that match his interests across the item popularity spectrum, while at the same time both suppliers have received exposure. This shows that the popularity calibration for the users would also benefit the suppliers. 

\section{Algorithmic Solutions}

The solutions for tackling popularity bias in the literature can be categorized into two groups:\footnote{A third option, preprocessing, is generally not useful for popularity bias mitigation because undersampling the popular items greatly increases the sparsity of the data.}

\begin{itemize}
    \item \textbf{Model-based:} In this group of solutions, the recommendation generation step is modified, so that the popularity of the items is taken into account in the rating prediction \cite{vargas2014improving,sun2019debiasing,abdollahpouri2017controlling,adamopoulos2014over}. 
    
    \item \textbf{Re-ranking:} A re-ranking algorithm takes a larger output recommendation list and re-orders the items in the list to extract a shorter final list with improved long-tail properties. Most of the solutions for tackling popularity bias fall into this category \cite{adomavicius2011maximizing,adomavicius2011improving,antikacioglu2017,flairs2019}. 
\end{itemize}

\subsection{Calibrated Popularity}

\label{sec:calibration}
Our proposed technique, \textit{Calibrated Popularity} (CP), is a re-ranking method. We build on the work of Steck~\cite{steck2018calibrated} in using re-ranking to provide results that better match the distributional properties of user profiles. In Steck's case, the distribution of interest was the distribution of genres across recommended movies. In our case, it is the distribution of the item popularity we seek to control\footnote{Note that unlike the genre labels in \cite{steck2018calibrated} where it is possible for a movie to have multiple genres, each item only belongs to one item group.}. 

CP algorithm operates on an initial recommendation list $\ell'$ of size $m$ generated by a base recommender to produce a final recommendation list $\ell$ of size $n$ ($m>>n$). Similar to \cite{steck2018calibrated}, we measure distributional differences in the categories (groups) to which items belong $C=\{c_1$,$c_2$,...,$c_k\}$. For our purposes, these are the three $H$, $M$ and $T$ item groups described above (i.e. $C=\{H,M,T\}$).

To do this comparison, we need to compute a discrete probability distribution $P$ for each user $u$, reflecting the popularity of the items found in their profile $\rho_u$ over each item group $c \in C$. We also need a corresponding distribution $Q$ over any given recommendation list $\ell$, indicating what item popularity groups are found among the listed items. For measuring the interest of each user towards each item popularity group, we use Vargas et al.'s \cite{vargas2013exploiting} measure of category propensity. Specifically, we calculate the propensity of each user $u$ towards each item group $c \in C$ in her profile $\rho_u$ ($p(c|u)$) and the ratio of such item group in her recommendation list $\ell_u$ ($q(c|u)$) as follows:  

\begin{equation}
p(c|u)=\frac{\sum_{i \in \rho_u}r(u,i)\mathbbm{1}(i \in c)}{\sum_{c_j \in C}\sum_{i \in \rho_u }r(u,i)\mathbbm{1}(i \in c_j)} \;, \;\;\;\;\; q(c|u)=\frac{\sum_{i \in \ell_u}\mathbbm{1}(i \in c)}{\sum_{c_j \in C}\sum_{i \in \ell_u }\mathbbm{1}(i \in c_j)}
\end{equation}   
$\mathbbm{1}(.)$ is the indicator function returning zero when its argument is False and 1 otherwise.

In order to determine if a recommendation list is calibrated to a given user, we need to measure the distance between the two probability distributions $P$ and $Q$. There are a number of metrics for measuring the statistical distance between two distributions~\cite{lin1991divergence}. Steck \cite{steck2018calibrated} used Kullbeck-Liebler (KL) Divergence. In this paper, we are using Jensen–Shannon divergence, which is a modification of KL Divergence that has two useful properties which KL divergence lacks: 1) it is symmetric: $\mathfrak{J}(P,Q)=\mathfrak{J}(Q,P)$ and 2) it has always a finite value even when there is a zero in $Q$. For our application, it is particularly important that the function be well-behaved at the zero point since it is possible for certain item groups to contain zero items in the recommendation list. Steck in \cite{steck2018calibrated} has fixed the problem at zero points by applying a smoothing technique but the Jensen–Shannon divergence already takes care of this issue.

Given the $KL$ as the KL divergence function, the Jensen–Shannon divergence ($\mathfrak{J}$) between two probability distributions $P$ and $Q$ is defined as follows:
\begin{equation}
    \mathfrak{J}(P,Q)=\frac{1}{2}KL(P,M)+\frac{1}{2}KL(Q,M), \; \; M=\frac{1}{2}(P+Q)
\end{equation}

Similar to \cite{steck2018calibrated,wasilewski2018intent}, we use a weighted sum of relevance and calibration for creating our re-ranked recommendations. In order to determine the optimal set $\ell^*$ from the $m$ recommended items, we use maximum marginal relevance:
\begin{equation}\label{mmr}
    \ell^*= \argmax_{\ell, |\ell|= n} (1-\lambda)\cdot Rel(\ell)-\lambda \mathfrak{J}(P,Q(\ell))
\end{equation}
\noindent where $\lambda$ is the weight controlling the relevance versus the popularity calibration and $Rel(\ell)$ is the sum of the predicted scores for items in $\ell$. Since smaller values for $\mathfrak{J}$ are desirable, we used its negative for our score calculation. 

As noted in \cite{steck2018calibrated}, finding the optimal set $\ell^*$ is an NP-hard problem. However, a simple greedy optimization approach is computationally effective. The greedy process starts with an empty set and iteratively adds one item from the larger list to the under-construction list until it reaches the desired length. At each step $j$, both $Rel(\ell)$ and $\mathfrak{J}(P,Q(\ell))$ are calculated using the union of the items that are already in the under-construction list and the item $i$ that is a candidate to be added to the list (i.e. $\ell@j \cup i$) and the item that gives the highest score will be added to the list. This greedy solution achieves a $(1-1/e)$ optimality to the best possible $\ell^*$ list with $e$ being Euler's number. 

\section{Experimental Methodology}

\subsection{Data}\label{data_section}
To incorporate our analysis of supplier-side fairness, we needed datasets where the item suppliers could be identified. We found two publicly available datasets for our experiments: the first one is a sample of the Last.fm (LFM-1b) dataset \cite{schedl2016lfm} used in \cite{dominik2019unfairness}. The dataset contains user interactions with songs (and the corresponding albums). We used the same methodology in \cite{dominik2019unfairness} to turn the interaction data into rating data using the frequency of the interactions with each item (more interactions with an item will result in higher rating). In addition, we used albums as the items to reduce the size and sparsity of the item dimension, therefore the recommendation task is to recommend albums to users. We considered the artists associated with each album as the supplier. We removed users with less than 20 ratings so only consider users for which we have enough data. The resulting dataset contains 274,707 ratings by 2,697 users to 6,006 albums. Total number of artists is 1,998. 
 
 The second dataset is the MovieLens 1M dataset \cite{movielens} \footnote{Our experiments showed similar results on MovieLens 20M, and so we continue to use MovieLens 1M for efficiency reasons.}. This dataset does not have the information about the suppliers. However, as we mentioned earlier, we considered the director of each movie as the supplier of that movie and we extracted that information from the IMDB API. Total number of ratings in the MovieLens 1M data is 1,000,209 given by 6,040 users to 3,706 movies. Overall, we were able to extract the director information for 3,043 movies reducing the ratings to 995,487. The total number of directors is 831. 
 
\subsection{Experimental Settings}

We used 80\% of each dataset as our training set and the other 20\% for the test. As with other re-ranking techniques, our method also needs a base algorithm to generate the initial list of recommendations for post-processing. We use \textit{Item-based Collaborative Filtering (Item-CF)} \cite{sarwar2001item} for this purpose and we call it \textit{Base} for the rest of the paper. This particular algorithm is well known to exhibit strong popularity bias amplification as we saw in Figure ~\ref{pop_amplify}, and so represents a particularly steep challenge for any re-ranking approach. Our experiments on $RankALS$ showed similar patterns but for space limitations we only report the results for \textit{Item-CF}. We intend to explore the interactions between $CP$ re-ranking and algorithm choice in our future work. 

Similar to \cite{kaya2019comparison} we set the size of the recommendations generated by the \textit{Base} algorithm to 100 ($m=100$), and the size of the final recommendation list is 10 ($n=10$).

For baseline comparisons, we used three recent state-of-the-art re-ranking techniques for mitigating popularity bias. These methods cover a spectrum of different designs including weighted combination methods, network flow, and rank merging. Note that the inclusion of these baselines is not to show how great our proposed technique is but rather to highlight the problems with optimizing for some of the existing metrics to control popularity bias in the literature as these baselines do. These baseline methods are described below.

\begin{itemize}
    \item \textbf{Discrepancy Minimization (DM)} \cite{antikacioglu2017}:
    In this method, the goal is to improve the total number of unique recommended items, also referred to as aggregate diversity (see Equation \ref{aggdiv}) of recommendations using minimum-cost network flow method to efficiently find recommendation sub-graphs that optimize diversity. Authors in this work define a target distribution of item exposure (i.e. the number of times each item should appear in the recommendations) as a constraint for their objective function. The goal is therefore to minimize the discrepancy of the recommendation frequency for each item and the target distribution. 
    
     \item \textbf{FA*IR (FS) \cite{zehlike2017}:} This method was originally used for improving group fairness in job recommendation and was adapted here to improve the fairness of recommendations in terms of head ($H$) versus long-tail ($M \cup T$) items in recommendations. The algorithm creates queues of protected and unprotected items and merges them using normalized scoring such that protected items get more exposure. We define protected and unprotected groups as long-tail and head items, respectively. We performed grid search over the two hyperparameters, proportion of protected candidates in the top $n$ items \footnote{Based on suggestion from the released code, the range should be in $[0.02,0.98]$} and significance level\footnote{Based on suggestion from the released code, the range should be in $[0.01,0.15]$}, using values of $\{0.25,0.5,0.75,0.95\}$ and $\{0.05,0.1,0.15\}$, respectively. 
     
    \item \textbf{Personalized Long-tail Promotion (XQ) \cite{flairs2019}:} In this method, inspired by the xQuAD algorithm for query result diversification \cite{santos2010exploiting}, the objective for a final recommendation list is a balanced ratio of popular and less popular (long-tail) items. We specifically included $XQ$ since, similar to our approach, it leverages user propensity towards popular items in its calculations and hence it can be categorized as an intent-aware \cite{wasilewski2018intent} long-tail promotion technique \cite{flairs2019}. However, its main focus is on a balanced distribution of popular versus non-popular items in the recommendation lists, and user propensity is not considered as a first priority but rather as a tie-breaker. Authors of this technique only defined two item categories: short-head and long-tail, with the short-head being $H$ and long-tail being $M \cup T$.
    
\end{itemize}

All three re-ranking baselines and our $CP$ method have hyperparameters that control the trade-off between relevance and a second criterion: diversification in $XQ$, fairness in $FS$, aggregate diversity in $DM$, and popularity calibration in $CP$. To establish a point of comparison across the algorithms, we varied these trade-off hyperparameters for each algorithm, and chose, for each, a hyperparameter that yields roughly equal recommendation precision. On MovieLens, this intersection point was $Precision\approx 0.21$, corresponding to approximately 0.01 (or 4.5\%) precision loss compared to the \textit{Base} precision of 0.22. For Last.fm, this point was $Precision\approx 0.122$, a very slight increase from the \textit{Base} precision of 0.12. 

We used LibRec \cite{guo2015librec} and \textit{librec-auto} \cite{mansoury2018automating} for running the algorithms.

\subsection{Evaluation}
A multistakeholder evaluation entails multiple metrics representing the perspective of different stakeholders on system performance. As we have selected the hyperparameters for each algorithm that yields the same accuracy, we do not include recommendation accuracy in our evaluation. Instead, we examine the behavior of these algorithms relative to different metrics of long-tail performance. Conventional metrics are those that look at a system's overall recommendation of long-tail items. We show these results, but in addition, we use metrics that are sensitive to the performance across different groups of stakeholders: suppliers in different popularity categories, items in different popularity categories, and users with different levels of interests in item popularity. 

\subsubsection{Overall}
To capture overall long-tail performance, a variety of metrics have been used in prior research. Let $L$ be the combined list of all recommendation lists given to different users (note that an item may appear multiple times in $L$, if it occurs in recommendation lists of more than one user). Let $I$ be the set of all items in the catalog and $U$ be the set of all users. We compute the following metrics:
\begin{itemize}
    \item \textbf{Aggregate Diversity:} The ratio of unique recommended items across all users:
    
    \begin{equation}\label{aggdiv}
       Agg\mbox{-}{Div}=\frac{\textbar \bigcup_{u \in U}\ell_u \textbar}{|I|} 
    \end{equation}
    
    Higher values for this metric indicate that the recommendations ``cover'' more of the item catalog.
    
     \item \textbf{Long-tail Coverage (LC):} The concept of long-tail coverage is used in many prior work in popularity bias \cite{adomavicius2011improving,flairs2019} and it measures the ratio of long-tail items (in our case it is $M \cup T$ since we have divided the long-tail into two separate groups) that appear in the recommendation lists of different users. Effectively, this is \textit{Agg-Div} applied only to the $M \cup T$ portion of the catalog.  
    \begin{equation}
         LC=\frac{\left| \bigcup_{u \in U}(\ell_u \cap (M \cup T))\right|}{|M \cup T|}
    \end{equation} 

    \item \textbf{Gini Index:} Measures the inequality across the frequency distribution of the recommended items. If one item is recommended frequently while other items are ignored, the Gini index will be high, therefore lower values for this metric are desirable. 
    
     \begin{equation}
        Gini(L)=1-\frac{1}{|I|-1} \sum_{k=1}^{|I|}(2k-|I|-1)p(i_k|L)
    \end{equation}
    \noindent where $p(i|L)$ is the ratio of occurrence of item $i$ is $L$.

\end{itemize}

\subsubsection{Supplier Groups}
Similar to \cite{mehrotra2018towards}, we operationalize the concept of fairness for the suppliers (i.e. artists and directors in Last.fm and MovieLens datasets, respectively) using their popularity. In \cite{mehrotra2018towards}, the authors grouped the artists in Spotify data into 10 equal-sized bins. As we noted in Section~\ref{intro}, we group the suppliers into three different bins based on their position in the popularity spectrum $S=\{S_1,S_2,S_3\}$ with the $S_1$ being the most popular suppliers followed by $S_2$ and $S_3$ that have lower popularity.
Following \cite{mehrotra2018towards}, we use \textit{Equity of Attention Supplier Fairness} (ESF), which operationalizes the idea that a supplier's chance of being recommended should be independent of the popularity bin to which they belong. The \textit{ESF} of a list of recommendations ($L$) given to all users from the suppliers' perspective is defined as:


\begin{equation}
     ESF=\sum_{i=1}^{|S|}\sqrt {\sum_{j \in L} \mathbbm{1}(A(j) \in S_i)}
 \end{equation}

\noindent where $S_i$ is the list of suppliers belonging to the popularity bin $i$, and $A(j)$ is a mapping function that returns the supplier of item $j$.

\textit{ESF} rewards sets that are diverse in terms of the different supplier bins represented, thus providing a fairer representation of different supplier bins. Given the nature of the function, there is more benefit to selecting suppliers from a bin not yet having one of its suppliers already chosen. When a supplier from a particular bin is represented in the recommendations, other suppliers from the same bin will be penalized due to the concavity of the square root function (e.g. $\sqrt{1}+\sqrt{1}>\sqrt{2}$). 

The \textit{ESF} measure does not, however, consider the inherent popularity of suppliers in different bins. We introduce an alternative metric that does so: \textit{Supplier Popularity Deviation} (SPD). 

For any supplier group $s$, $SPD(s)=q(s)-p(s)$ where $q(s)$ is the ratio of recommendations that come from items of supplier group $s$ (i.e. $q(s)=\frac{\sum_{u \in U}\sum_{j \in \ell_u}\mathbbm{1}(A(j) \in s)}{n\times|U|}$), and $p(s)$ is the ratio of ratings that come from items of supplier group $s$ (i.e. $p(s)=\frac{\sum_{u \in U}\sum_{j \in \rho_u}\mathbbm{1}(A(j) \in s)}{\sum_{u \in U} |\rho_u|}$). The average \textit{SPD} across different groups of suppliers can be calculated as:

\begin{equation}
SPD=\frac{\sum_{i=1}^{|S|}|SPD(S_i)|}{|S|}
\end{equation}

Lower values for $SPD$ indicate a better match between the distribution of items from different suppliers in rating data and in the recommendations. The inverse $1-SPD$ can be considered a type of proportional fairness metric since it measures how the items from different supplier groups are exposed to different users \textit{proportional} to their popularity in rating data. 

\subsubsection{Item Groups}
We can use the same formalization as \textit{SPD} to define a metric \textit{Item Popularity Deviation} (IPD) that looks at groups of items across the popularity distribution. For any item group $c$, $IPD(c)=q(c)-p(c)$ where $q(c)$ is the ratio of recommendations that come from item group $c$ (i.e. $ q(c)=\frac{\sum_{u \in U}\sum_{j \in \ell_u}\mathbbm{1}(j \in c)}{n\times|U|}$). $p(c)$ is the ratio of ratings that come from item group $c$ (i.e. $p(c)=\frac{\sum_{u \in U}\sum_{j \in \rho_u}\mathbbm{1}(j \in c)}{\sum_{u \in U} |\rho_u|}$). The average \textit{IPD} across different groups of items can be measured as:

\begin{equation}
IPD=\frac{\sum_{c \in C}|IPD(c)|}{|C|},
\end{equation}


\subsubsection{User Groups}
Finally, based on our finding above that users with different interests in popular items may receive results with different degrees of calibration, we can examine the popularity calibration of recommendation results across groups of users with different popularity propensity. This metric is \textit{User Popularity Deviation} (UPD). For any user group $\textsl{g}$, $UPD(\textsl{g})=\frac{\sum_{u \in \textsl{g}}\mathfrak{J}(P(\rho_u),Q(\ell_u))}{|g|}$. The average \textit{UPD} across different groups of users is:
\begin{equation}
    UPD=\frac{\sum_{g \in G}UPD(\textsl{g})}{|G|}
\end{equation}
$UPD$ can be also seen as the average popularity miscalibration of the recommendations from the perspective of users in different groups.

For all three metrics \textit{IPD}, \textit{SPD} and \textit{UPD}, lower values are desirable. 

\section{Results and Discussion}
In this section we discuss the performance of different algorithms relative to these evaluation metrics.

\begin{table}[]
\large
\captionof{table}{Results of different algorithms on two datasets. The setting that yields the same precision for all four re-ranking methods $XQ$, $DM$, $FS$, and $CP$ is chosen and the results are reported. Bold values are statistically significant compared to the second best value in each column with significance level of 0.05 ($\alpha$=0.05). Up arrows indicate larger values are desirable while down arrows indicate smaller values are better. }\label{tab:overall}

\begin{tabular}{lcccccccc}
                                                & \multicolumn{1}{l}{}      & Agg-Div$\uparrow$ & LC$\uparrow$   & Gini$\downarrow$ & ESF$\uparrow$  & IPD$\downarrow$ & UPD$\downarrow$ & SPD$\downarrow$ \\ \bottomrule
\multicolumn{1}{l|}{\multirow{5}{*}{MovieLens}} & \multicolumn{1}{c|}{Base} & 0.029             & 0.003          & 0.950            & 2.349          & 0.534           & 0.610           & 0.475           \\
\multicolumn{1}{l|}{}                           & \multicolumn{1}{c|}{XQ}   & 0.107             & 0.0836         & 0.888            & 2.524          & 0.333           & 0.485           & 0.326           \\
\multicolumn{1}{l|}{}                           & \multicolumn{1}{c|}{DM}   & \textbf{0.127}    & \textbf{0.107} & \textbf{0.734}   & 2.630          & 0.343           & 0.478           & 0.278           \\
\multicolumn{1}{l|}{}                           & \multicolumn{1}{c|}{FS}   & 0.07              & 0.054          & 0.904            & 2.985          & 0.200           & 0.301           & 0.226           \\
\multicolumn{1}{l|}{}                           & \multicolumn{1}{c|}{CP}   & 0.076             & 0.062          & 0.900            & \textbf{3.160} & \textbf{0.133}  & \textbf{0.233}  & \textbf{0.128}  \\ \hline
\multicolumn{1}{l|}{\multirow{5}{*}{Last.fm}}   & \multicolumn{1}{c|}{Base} & 0.131             & 0.089          & 0.942            & 3.658          & 0.470           & 0.594           & 0.350           \\
\multicolumn{1}{l|}{}                           & \multicolumn{1}{c|}{XQ}   & 0.206             & 0.165          & 0.917            & 3.834          & 0.395           & 0.514           & 0.311           \\
\multicolumn{1}{l|}{}                           & \multicolumn{1}{c|}{DM}   & \textbf{0.393}    & \textbf{0.356} & \textbf{0.749}   & 3.957          & 0.313           & 0.459           & 0.242           \\
\multicolumn{1}{l|}{}                           & \multicolumn{1}{c|}{FS}   & 0.212             & 0.179          & 0.867            & 4.091          & 0.203           & 0.374           & 0.225           \\
\multicolumn{1}{l|}{}                           & \multicolumn{1}{c|}{CP}   & 0.260             & 0.233          & 0.831            & \textbf{4.240} & \textbf{0.119}  & \textbf{0.281}  & \textbf{0.168}  \\ \bottomrule
\end{tabular}
\end{table}
\subsection{Overall performance}
The overall results for different algorithms on both MovieLens and Last.fm datasets can be seen in Table ~\ref{tab:overall}. The arrow next to the metric name indicates the direction of preference for a better outcome. 

The three overall metrics, aggregate diversity (\textit{Agg-Div}), long-tail coverage ($LC$), and Gini index (\textit{Gini}), all show similar patterns on both datasets. On MovieLens, the $DM$ algorithm has the best performance on these metrics with $XQ$ next, and $FS$ and $CP$ behind. Not only is greater coverage achieved for the whole catalog, but a better distribution as shown by the Gini index. For the Last.fm dataset, $DM$ still shows the most improvement, but on this denser dataset $XQ$ does not perform as well as $FS$, and $CP$ comes in second throughout. 

\begin{figure*}
\centering
\SetFigLayout{3}{2}
 \subfigure[MovieLens]{\includegraphics[width=5.9in]{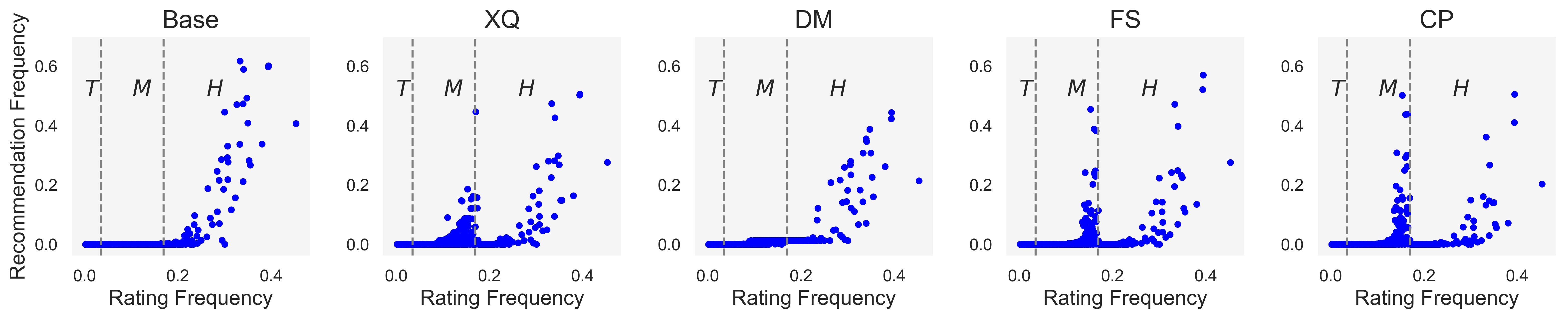}}
\subfigure[Last.fm]{\includegraphics[width=5.9in]{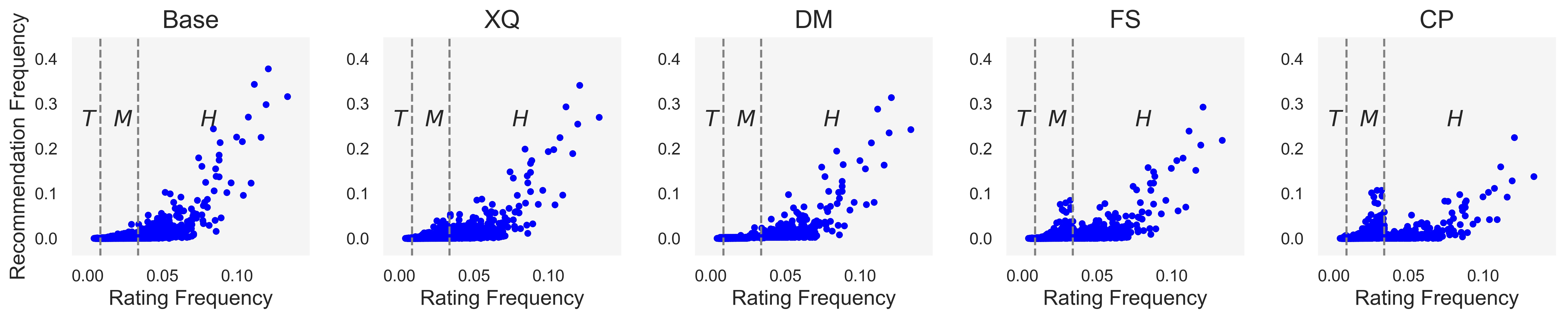}}
\caption{Exposure rate vs popularity across different re-ranking algorithms} \label{expoure_scatter}
\end{figure*}

A more detailed picture of the way that item popularity is transformed in recommendation frequency is available in Figure~\ref{expoure_scatter}. These scatter plots have rating frequency and popularity segments (i.e. item groups) on the x-axis with the \textit{Tail} items ($T$) having low rating frequency on the left, moving to the popular \textit{Head} items ($H$) on the right. Also, y-axis shows the frequency of the recommendations for each item (i.e. exposure rate). Looking first at part (a) showing the MovieLens results, we see that, in the \textit{Base} algorithm, the $T$ and $M$ items are rarely recommended, as previously noted. The $XQ$, $FS$, and $CP$ algorithms all boost the $M$ items significantly, while a more limited effect is seen for $DM$. The division of the items into groups has an obvious effect as the promotion of lower-popularity items is concentrated at the top end of the $M$ items, which could be expected as these are the items in this set with the most number of ratings. Note also that all algorithms reduce the number of items with extreme rating frequency (contained in > 50\% of recommendation lists). There are a handful of these items in the \textit{Base} recommender, and still two such items in $FS$, but none for any of the others. 

In part (b) of the figure, we see the results for Last.fm. The \textit{Base} recommender is not quite as extreme in its neglect of the lower parts of the popularity distribution here and that gives all of the re-rankers more to work with. All of the algorithms boost the frequency of the recommendations for $M$ items somewhat and reduce the extremes of recommendation frequency. However, this effect is most pronounced in the $CP$ algorithm.

This figure helps explain the performance of the $DM$ algorithm on the \textit{Agg-Div}, \textit{LC}, and \textit{Gini} metrics. On the MovieLens dataset, the $DM$ algorithm promotes some of the $M$ items as can be seen in the small step increase in the plot in the $M$ region. Thus, it increases aggregate and long-tail coverage since in both of these two metrics, the recommendation frequency does not matter meaning recommending an item even once would be counted the same as recommending it to many users. By giving very small but consistent exposure to this range of items, it improves the Gini index as well.

\subsection{Multistakeholder Performance}

Without a multistakeholder analysis, we might conclude from the overall metrics shown in Table~\ref{tab:overall} that $DM$ is a superior approach for remediating popularity bias. However, evaluating the recommendations from the perspective of the different stakeholders reveal a different story. Because $CP$ aims to improve popularity calibration, we might expect that \textit{UPD}, which more or less measures the inverse of average calibration, would be improved and this is indeed a strong effect across both datasets. $CP$ is applying its long-tail enhancement where it matters for users and therefore is able to have a big impact on calibration. 
Thus, we might say that \textit{UPD} is measuring \textit{useful long-tail diversity} in recommendation results. Interestingly, this emphasis on users also has a beneficial effect on the metrics related to other recommendation stakeholders as shown in the superior performance on \textit{ESF}, \textit{IPD}, and \textit{SPD}. 

\begin{figure*}
\centering
\SetFigLayout{3}{2}
 \subfigure[MovieLens]{\includegraphics[width=5.9in]{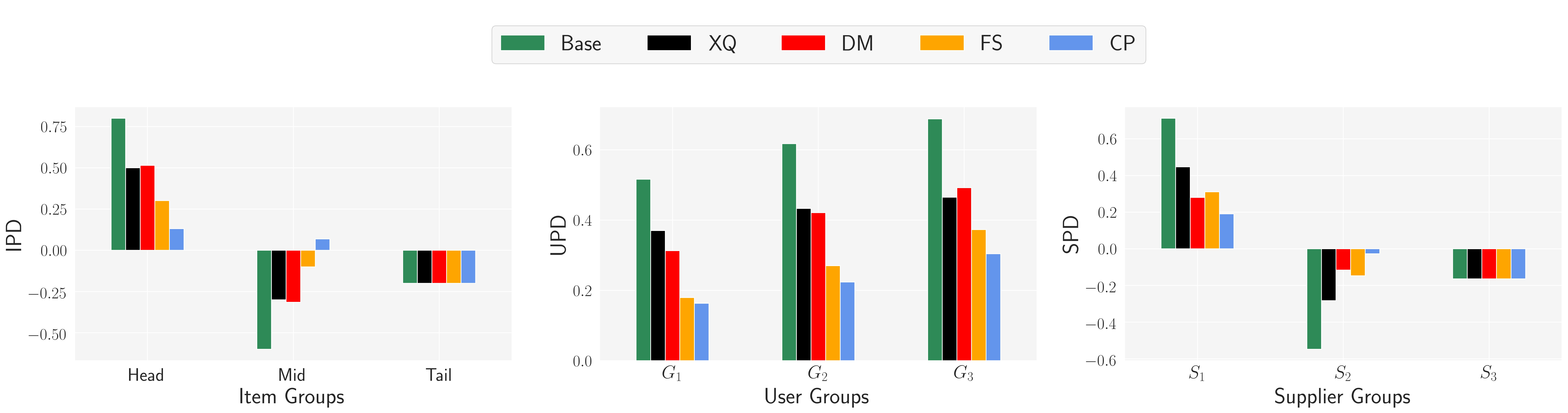}}
\subfigure[Last.fm]{\includegraphics[width=5.9in]{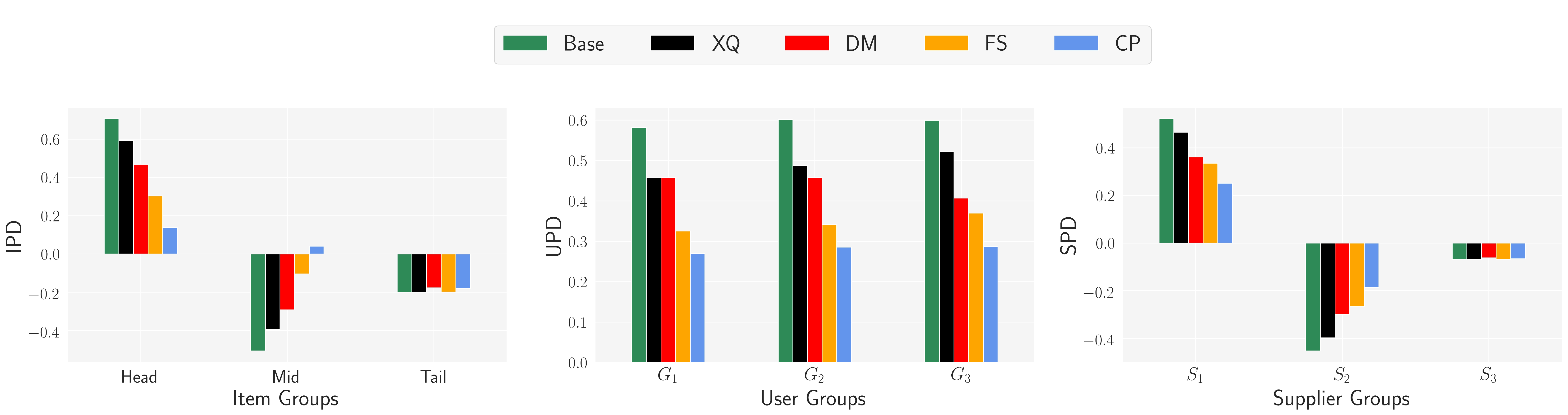}}

\caption{Popularity deviation metrics for different stakeholder groups} \label{groups_analysis_deviation}
\end{figure*}

Additional detail on the comparative performance for different stakeholders and sub-groups can be seen in Figure~\ref{groups_analysis_deviation}. These results compare the popularity deviation metrics (\textit{IPD}, \textit{UPD}, \textit{SPD}) for different sub-groups. 

\subsubsection{Item Groups}
First, we can see that all algorithms have positive \textit{IPD} values for items in \textit{Head} meaning they all over-recommend popular items. This is not surprising since the \textit{Base} algorithm exhibits a strong bias towards these items and although the re-ranking techniques have reduced this over-concentration they still have not removed it completely. The $CP$ algorithm comes the closest to doing so. Also, in all algorithms except for $CP$, items from the \textit{Mid} group have received negative deviation meaning they are under-represented in the recommendations. But again, $CP$ performs significantly better in the \textit{Mid} group giving these items a slight boost, as we also saw in Figure~\ref{expoure_scatter}. 

The result for the \textit{Tail}, however, show that even though $CP$ has slightly lower deviation for this item group on Last.fm (not significant) compared to other algorithms, all algorithms including $CP$ have performed poorly. This is also consistent with what we found in Figure~\ref{expoure_scatter}. Our analysis showed that even the original list ($m=100$) did not contain many items from the \textit{Tail} group, hence, it was impossible for the re-ranking approaches to perform any better. We also tried a larger initial list size ($m=500$) and we noticed the performance for the \textit{Tail} items does increase but, at the same time, the precision drops. 
\subsubsection{User Groups}
Looking at the charts for user groups, we, again, see the superiority of $CP$. Firstly, $G_1$, the group with the highest interest towards popular items, has the lowest \textit{UPD} (i.e. lowest miscalibration) using all algorithms and $G_3$ has the highest miscalibration. That shows users with high interest towards popular items are served much better as expected. Regardless, all user groups have experienced the lowest miscalibration using $CP$. $FS$ also performs well here with lower \textit{UPD} compared to $DM$ and $XQ$. What is interesting about $CP$ and $FS$ with regard to \textit{UPD} is that they both have comparable \textit{UPD} for $G_1$ but the \textit{UPD} of $CP$ is significantly lower for $G_2$ and $G_3$ indicating our algorithm also performs well for users with lesser interest in popular items as it calibrates their recommendation better. The same for $DM$ and $XQ$ where $DM$ works better for users in $G_1$ (lower \textit{UPD}) while $XQ$ is superior for $G_3$. These are distinctions that we would not have observed without a multistakeholder approach to evaluation.

\subsubsection{Supplier Groups}\label{supplier_groups}

Finally, the chart for supplier groups similarly confirms the superiority of $CP$ in this multistakeholder analysis. On both datasets, $CP$ has the lowest \textit{SPD} for the $S_1$ (group with high popularity) and $S_2$ (group with medium popularity) groups. On Movielens, this metric for $S_2$ is close to zero meaning $CP$ has given a fair chance of being exposed to suppliers from this group. Similar to item groups, all algorithms have performed poorly on $S_3$ (the suppliers with lowest popularity). 

One other interesting finding that can be seen in this chart for MovieLens dataset is the relationship between item popularity and supplier popularity. One might expect that when an algorithm performs better than another on items with high popularity ($Head$) it should also perform better on the suppliers with high popularity as these two seem to be correlated. However, as we can see, $FS$ performs better than $DM$ on the \textit{Head} items (lower $IPD$) but its performance on $S_1$ is worse (higher $SPD$). The reason is, we observed not every item from a popular supplier is necessarily popular and therefore it is possible for an algorithm to perform differently on item groups and supplier groups as it is the case on MovieLens dataset. 

\begin{figure}
    \centering
    \includegraphics[width=6in]{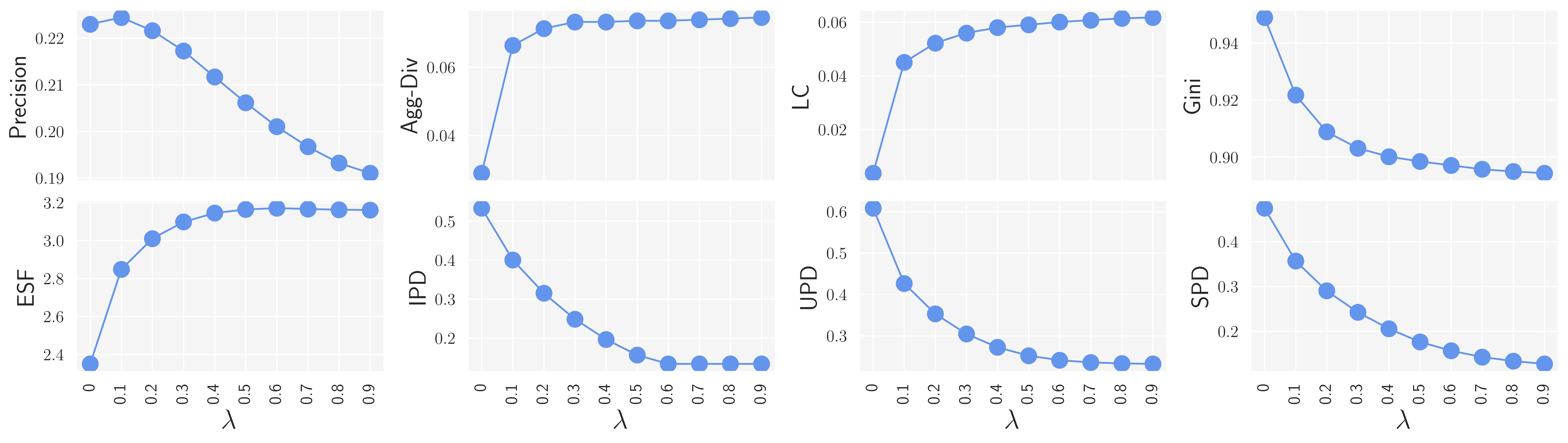}
    \caption{The effect of $\lambda$ in the $CP$ algorithm (MovieLens)}
    \label{lambda_analysis:movielens}
\end{figure}
\begin{figure}
    \centering
    \includegraphics[width=6in]{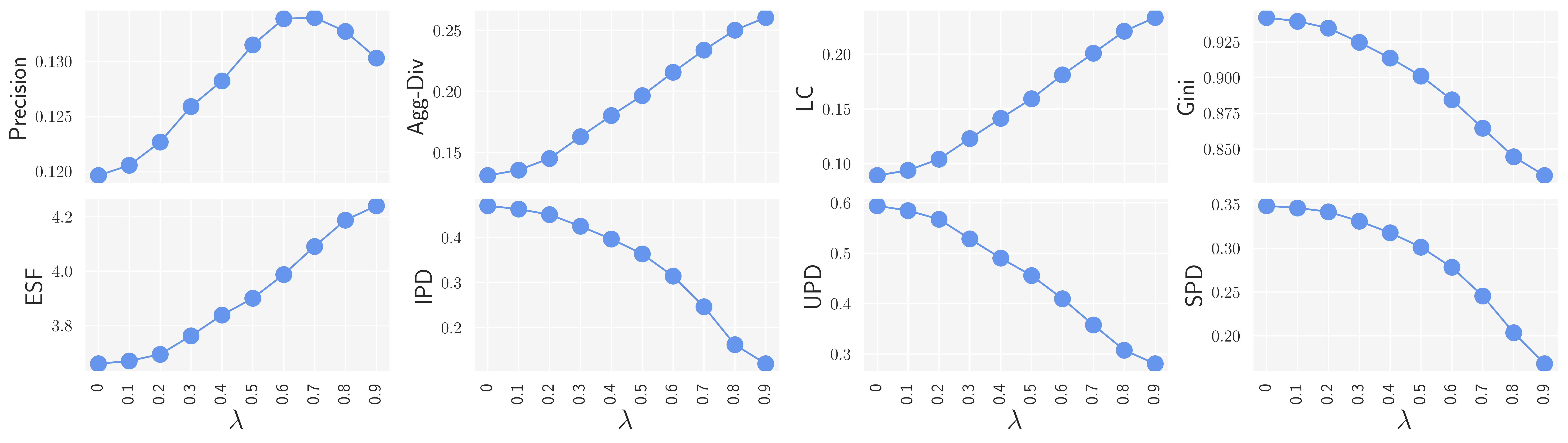}
    \caption{The effect of $\lambda$ in the $CP$ algorithm (last.fm)}
    \label{lambda_analysis:lastfm}
\end{figure}

\subsection{Sensitivity Analysis}
The previous analysis depended on versions of the re-ranking algorithms that had approximately equal precision. Figures ~\ref{lambda_analysis:movielens} and ~\ref{lambda_analysis:lastfm} show the performance of the $CP$ algorithm for different values of $\lambda$ in Equation ~\ref{mmr}, showing the interaction between this hyperparameter and our evaluation metrics. On the MovieLens data, giving more weight to the calibration component in Equation ~\ref{mmr} (higher values for $\lambda$) causes a drop in precision: a small peak at $\lambda = 0.1$ and roughly 0.03 loss when $lambda=0.9$. All other metrics are improving as $\lambda$ increases. Looking at $\lambda=0.2$ we can see that with a negligible loss in precision (approximately $0.001$) all other metrics have significantly improved. 

On Last.fm the results are even more promising. The precision values over different $\lambda$ values are slightly improved using our CP method\footnote{Comparing the y-axis scale of the corresponding plots for both datasets, we can see that the precision improvement on Last.fm is not that high.}, therefore even with the highest $\lambda$ there is no loss in precision while all other metrics have significantly improved. One key difference between the datasets is that the aggregate diversity for the \textit{Base} algorithm on Last.fm is much higher. This gives the re-ranking algorithms much more to work with as they attempt to include diverse items.

\section{Conclusion and Future Work}

In this work, we have taken a multistakeholder view of the problem of popularity bias. We show that large segments of the user population in a typical recommender system have a strong interest in items outside of the ``short head'' of the distribution. Consistent with prior work, we show that because these long-tail items are recommended less frequently, these groups of users are not well-served and that this problem has a ripple-out effect on the other recommendation stakeholders. This multistakeholder evaluation of popularity bias has revealed certain aspects of the algorithms' behavior that cannot be captured using standard evaluation methods and metrics that are focused on overall outcomes, including differences across results returned to different user, supplier, and item groups. 

We define an algorithm for improving the popularity calibration of recommendation outputs and demonstrate its superiority over approaches that merely attempt to reduce popularity bias without considering calibration. We show that this approach also improves supplier fairness, as measured by the exposure of items from different suppliers in recommendation lists using two different definitions for measuring supplier fairness.

A number of questions remain for future work. Any re-ranking algorithm is, by definition, independent of the base algorithm that generates the results. In our work here, we have used the item-based collaborative filtering algorithm, which is well-known to have strong popularity bias amplification and that was one of the reasons we picked it as our base algorithm. We have seen that there is a limit to how many \textit{Tail} items can be re-ranked, because the algorithm simply does not return them. The trade-offs between coverage, popularity calibration, and precision explored here may be different for different base algorithms and such effects will be important to explore in future work.

The beneficial side-effect of supplier fairness was not directly optimized as our algorithm focused on the user calibration and no information about the suppliers was used in the calibration algorithm. However, because of the positive correlation between item popularity and the average popularity of the corresponding suppliers, improving popularity calibration has this indirect effect of improving supplier fairness. An interesting extension of this work would be to incorporate the tendency of each user towards different groups of suppliers as well as different item popularity categories in the calibration algorithm. We will leave this for future work. 

The interaction between content category calibration (as the movie genres used in \cite{steck2018calibrated}) and popularity calibration as developed here is an open question. Obviously, some genres are more popular than others, but our preliminary work has shown that optimizing for genre calibration is not sufficient to achieve popularity calibration. Combining multiple types of calibration is an interesting future challenge. 

It is also an open question how users experience popularity calibration / miscalibration in recommendation lists. A user study that varied the calibration of results for users in different popularity preference categories would help quantify this factor, compared to others that influence user receptivity to recommendation results.

\bibliographystyle{ACM-Reference-Format}
\bibliography{sample-base}
\end{document}